%% file: main.tex
\documentclass[conference]{IEEEtran}
\IEEEoverridecommandlockouts
\usepackage{amsmath,amssymb,amsfonts}
\usepackage{algorithmic}
\usepackage{graphicx}
\usepackage{textcomp}
\usepackage{xcolor}
\usepackage[ruled, linesnumbered]{algorithm2e}
\usepackage{textcomp}
\usepackage{booktabs}
\usepackage[numbers,square,comma,compress,sort]{natbib}
\def\BibTeX{{\rm B\kern-.05em{\sc i\kern-.025em b}\kern-.08em
    T\kern-.1667em\lower.7ex\hbox{E}\kern-.125emX}}

\begin{document}

\title{\vspace{-1mm}Online Adaptive Learning for Runtime Resource Management of Heterogeneous SoCs\vspace{-4.25mm}}
\author{Sumit K. Mandal$^1$, Umit Y. Ogras$^1$, Janardhan Rao Doppa$^3$, Raid Z. Ayoub$^2$, Michael Kishinevsky$^2$, Partha P. Pande$^3$ \\
$^1$ School of Electrical, Computer and Energy Engineering, Arizona State University\\
$^2$ Strategic CAD Labs, Intel Corporation, Hillsboro, OR\\
$^3$ School of Electrical Engineering and Computer Science, Washington State University, Pullman, WA
\vspace{-3.25mm}}

\maketitle


\begin{abstract}
Dynamic resource management has become one of the major areas of research in modern computer and communication system design 
due to lower power consumption and higher performance demands.
The number of integrated cores, level of heterogeneity and amount of control knobs increase steadily.
As a result, the system complexity
is increasing faster than our ability to optimize and dynamically manage the resources. 
Moreover, offline approaches are sub-optimal due to workload variations and large volume of new applications unknown at design time.
This paper first reviews recent online learning techniques for predicting system performance, power, and temperature.
Then, we describe the use of predictive models for online control using two modern approaches: imitation learning (IL) and an explicit nonlinear model predictive control (NMPC). 
Evaluations on a commercial mobile platform with 16 benchmarks show that the IL approach successfully adapts the control policy to unknown applications.
The explicit NMPC provides 25\% energy savings compared to a state-of-the-art algorithm for multi-variable power management of modern GPU sub-systems.

\end{abstract}

\input{files/introduction.tex}

\input{files/overview_background.tex}
\input{files/perf_modeling.tex}
\input{files/control.tex}
\input{files/conclusion.tex}

\bibliographystyle{ieeetr}
\footnotesize{
\bibliography{main.bbl}
}

\end{document}

%% file: files/introduction.tex
\section{Introduction}




Heterogeneous systems-on-chip (SoCs) have become the backbone for a multitude of
electronic devices ranging from handheld smartphones~\cite{ODROID_Platforms} to high-performance computers~\cite{jeffers2016intel}.
They typically integrate multiple CPU cores, GPU, and specialized accelerators, such as 
digital signal processor (DSP), vector processing unit, 
image processing unit, 
and wireless communication modems.
While the increasing number of processing elements 
improves device performance, 
it also poses a challenge for dynamic resource management (DRM) of on-chip resources. 
This challenge is exacerbated with dynamic variations in the workloads and new applications that introduce unexpected features
not known at the design time.
For example, the optimal use of CPU and specialized processor resources, such as the number of active cores and their power states, 
vary both as a function of active applications at a coarse scale (in the order of seconds to minutes) 
and their rapidly varying requirements (in the order of microseconds to milliseconds).
Therefore, there is need for more research 
on the broad topic of runtime resource management of modern SoCs.

Different processing elements in heterogeneous SoCs can operate at multiple voltage and frequency levels.
For example, Samsung Exynos 5422 SoC with big.LITTLE architecture supports 4940 unique configurations of control variables~\cite{gupta2017dypo}. 
Similarly, the state space is enormous for SoCs used in server systems.
For example, Xeon Phi\texttrademark\  has 72 cores and each core can 
run at many different frequency levels~\cite{jeffers2016intel}. 
Even with two frequency levels, 
this Xeon Phi system offers more than 4$\times$10$^{21}$ combinations that can be chosen dynamically.
The optimal use of these control variables
depends on the workload and the optimization objective, such as power, performance, and performance-per-watt.
To shorten the control feedback, a significant part of the DRM in modern SoCs runs in firmware on on-die micro-controllers, which have both limited compute
and memory resources for low power operation. Operating system power management drivers are also resource constrained. Hence, the frequency governors embedded in operating systems employ simple techniques to manage resources. 
For example, interactive and on-demand governors increase (or decrease) operating frequency of cores when the utilization of the cores goes above (or below) a predefined threshold~\cite{pallipadi2006ondemand}.
These heuristics leave considerable room for improvement in both performance and power consumption.
Thus, there is a strong need for techniques that can select the optimal configuration to trade-off between performance and power consumption.

In the past, researchers proposed SoC resource management techniques designed offline.
Some of these techniques are specific to mobile processors~\cite{gupta2017dypo, vallina2012energy, pathania2014integrated}, while others are applicable to SoCs in general~\cite{kanduri2018approximation, moazzemi2019hessle, ogras2013modeling}.
These techniques profile the target heterogeneous platform by changing the configuration of the control
variables of the SoC~\cite{pasricha2020survey}.
Then, a controller is constructed to optimize a target objective by using the data obtained from profiling.
However, offline techniques are effective only if all potential applications are known at the design-time. With the current prolific growth of applications in smartphones, personal computers, and data-centers, it is impractical to assume that all applications are known while designing the SoC for most of the application domains.
Therefore offline resource management techniques are not guaranteed to perform well for new applications at runtime. 

\begin{figure*}[t]
	\centering
	\vspace{-2mm}
	\includegraphics[width=0.78\linewidth]{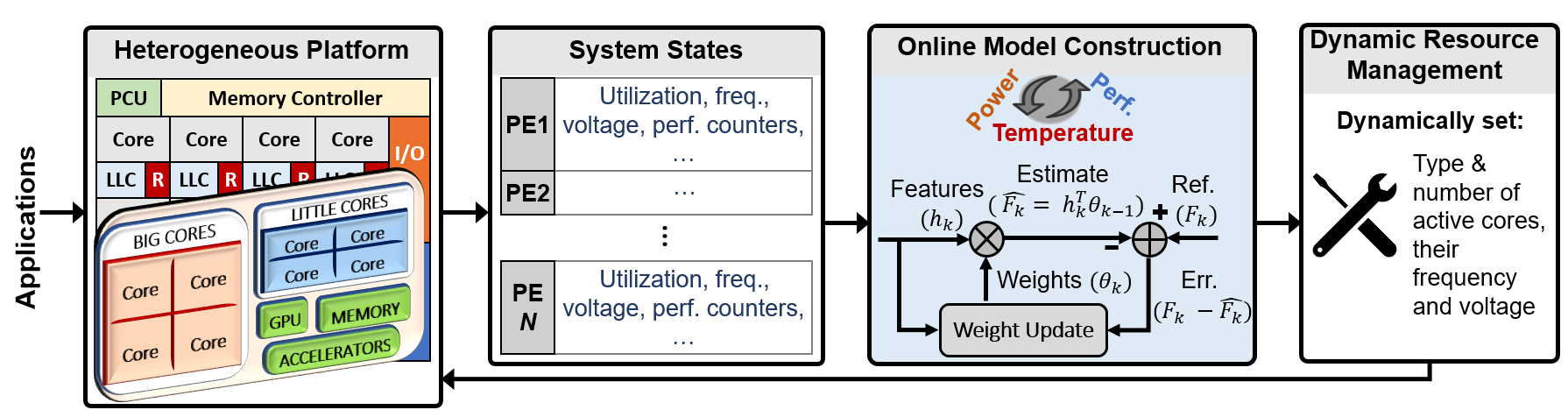}
	\vspace{-3mm} 
	\caption{Online learning framework for modeling design objectives and model-guided dynamic resource management methodology. Online adaptive learning approaches presented in Section~\ref{sec:ana_model} construct analytical models of key design metrics including performance, power consumption, and temperature. 
	DRM techniques discussed in Section~\ref{sec:control} employ these models to manage the set of active processing elements and their power states.}
	\vspace{-5mm}
	\label{fig:overview}
\end{figure*}

Control-theoretic approaches are also proposed for modeling and management of resources in SoCs.
For example, an offline auto-regressive performance model for GPUs is proposed in~\cite{dietrich2014lightweight}. However, offline performance models are not guaranteed to deliver expected efficiency while executing new applications online. 
To overcome this difficulty, recursive least square-based online learning technique to model GPU performance is proposed in~\cite{gupta2018online}. 
Similarly, online learning techniques have been proposed for dynamic resource management of heterogeneous processors~\cite{mandal2020energy}. 
A recent online resource management technique uses deep-Q network based reinforcement learning (RL) approach~\cite{zhang2017energy}.
RL-based techniques have notable drawbacks: designing the reward function is not trivial and requires a large data-set due to the trial-and-error learning process to converge to a near-optimal policy.
Hence, they are not practical for SoC resource management.

Imitation learning (IL) is more suitable for resource management compared to RL, since it requires fewer data samples to converge to an optimal policy~\cite{schaal1999imitation,IL1,IL2}. 
Furthermore, it has lower implementation complexity,
which allows for software implementation, e.g., as part of the OS governor.
In this methodology, first a control policy, known as the Oracle, is constructed offline.
Since the Oracle requires large storage, IL methods 
constructs a policy that approximates the Oracle using supervised learning techniques. 
Several studies applied this methodology offline to construct 
dynamic resource management policies for different hardware platforms~\cite{mandal2019dynamic, kim2017imitation}. 
However, offline Oracles are limited by the training data available at design time. Hence, they may not generalize to new types of workloads which are unknown at design time. 
This challenge has recently been addressed 
by an online-IL approach that combines offline and online optimization~\cite{mandal2020energy}. 
First, the workload information available at design time is used offline to construct an IL policy, as well as analytical power and performance models, similar to the previous approaches~\cite{mandal2019dynamic, kim2017imitation}. 
This policy and models serve as the initial points at runtime.
Then, the online power and performance models are used to evaluate the IL policy decisions with respect to candidate configurations that could have been chosen. 
If the IL-based policy fails to select the best configuration among the available candidates,
then the policy is updated at runtime using the supervision given by the analytical models. 
Hence, the IL policy adapts to new workloads at runtime~\cite{mandal2020energy}.



Model predictive control (MPC) has recently been applied to 
dynamic resource management of SoCs~\cite{chakrabarty_2017, zhang_2016, mercati_2017}.
MPC constructs a non-linear constrained optimization problem, that can efficiently be solved only offline. 
To enable online application, explicit model predictive control technique approximates the surface of MPC using simple machine learning regression models that can be implemented at low overhead while achieving near optimal control. 
This technique uses not only power and performance models (that are constructed offline and can be adapted online), but also models of sensitivity of optimization objectives (power and performance) to changes in control variables, 
such as frequency and the number of active cores. 
These sensitivity models 
can effectively adapt the control policy to a particular application, even if the core control algorithm remains the same. 
Hence, this technique, called explicit non-linear MPC (ENMPC), 
can be implemented with a low memory and runtime complexity that suits  
hardware and firmware implementations. 


%% file: files/overview_background.tex
\vspace{-1mm}
\section{Overview of Online Learning Framework}
\vspace{-1mm}

This paper presents a short overview of online
learning methods for runtime resource management of SoCs, as illustrated in Figure~\ref{fig:overview}.
The goal of this framework is two-fold: analytical modeling of design metrics and model-guided learning of resource management policies.

The modeling aspect aims to characterize the key design metrics, such as power consumption, performance, and temperature, as a function of both application workload and system parameters. 
SoCs are equipped with multiple on-chip performance counters and sensors which at runtime provide the information on system states, such as utilization of the CPU cores, number of retired instructions, number of cache misses and memory bandwidth, 
and temperature.
It is possible to construct analytical SoC power and performance models as functions of these system states.
Initial models can be constructed offline using design-time information.
Then, they can adapt to time-varying workload characteristics through light-weight online learning algorithms, as shown in Figure~\ref{fig:overview}. 
It is also possible to build similar models for the sensitivity of control objectives to input  variables, i.e., derivatives of the above functions with respect to control variables that can predict the effect of the control decision on the future iterations. 
As representative cases, this paper presents power and temperature modeling approaches of SoCs (Section~\ref{sec:power_model}), adaptive performance modeling of integrated GPUs (Section~\ref{sec:gpu_model}), and performance modeling of networks-on-chip (NoC) (Section~\ref{sec:noc_model}).

Model-guided policy learning approaches utilize design-time knowledge of the target architecture and analytical models to effectively learn 
DRM policies that optimize the trade-off between design objectives such as power and performance. Design-time knowledge includes the number and types of processing elements in the target SoC, as well as their power states that can be controlled at runtime. 
This information can be used with analytical models and experimental measurements to construct an Oracle that maps different hardware configurations and power states to key performance metrics in the IL approach, or it can be used to predict and adapt the predictive sensitivity functions of the ENMPC approach, in both cases leading to adapting the control decisions. Both techniques are discussed in Section~\ref{sec:control}. 

%% file: files/perf_modeling.tex
\section{Learning Runtime Analytical Models} \label{sec:ana_model}





\subsection{Power and Thermal Modeling For CPUs} \label{sec:power_model}

Power and thermal models are essential parts of SoC resource management techniques.
A power modeling technique for modern heterogeneous SoCs are presented in~\cite{brooks2007power}.
A temperature estimation methodology for mobile platforms is proposed in~\cite{bhat2017algorithmic}.
This methodology can be used both for predicting the temperature of thermal hotspots at a future time instant and for computing the maximum power consumption that can be sustained before causing thermal violations. 
Then, the power budget is used as a metric to throttle the frequency and number of operating cores to avoid temperature violations. 
Similarly, an analysis of power-temperature stability and safety of mobile SoCs is presented in~\cite{bhat2017power}.
This work analyzes the power-temperature dynamics of the target
SoC. It derives the necessary and sufficient conditions for the
existence and stability of the thermal fixed point, 
which is defined as the steady state temperature reached under a given average power consumption.
It also presents a methodology to find the thermal fixed points of a given SoC at runtime as a function of the power consumption of its active cores.


Thermal hotspots are no longer limited to die and package temperatures in mobile systems. Skin temperature has emerged as a new class of hot spots which impacts user experience as it may cause discomfort and harm human skin. It should be kept within the desired limits using closed-loop thermal management algorithms. 
One big challenge is providing skin temperature feedback to the controller since its direct measurement is not feasible in practice. The solution is to estimate skin temperature using models that take input from internal sensors. 
It has been shown that machine learning models for skin temperature estimation can be coupled with temperature-aware dynamic voltage and frequency scaling (DVFS) algorithms~\cite{Egilmez_skin_temp}. 
Similarly, a recent technique incorporates an online learning technique to model and manage the skin temperature of mobile SoCs~\cite{chetoui2020coordinated}. 
Finally, skin temperature estimation can be enhanced using algorithms for sensor selection~\cite{zhang_sensor_placement}, that improves the placement of internal sensors.



\vspace{-0.5mm}
\subsection{Online Performance Modeling for Integrated Cores} \label{sec:gpu_model}
\vspace{-0.5mm}

\begin{figure}[t]
	\centering
	\includegraphics[width=0.9\linewidth]{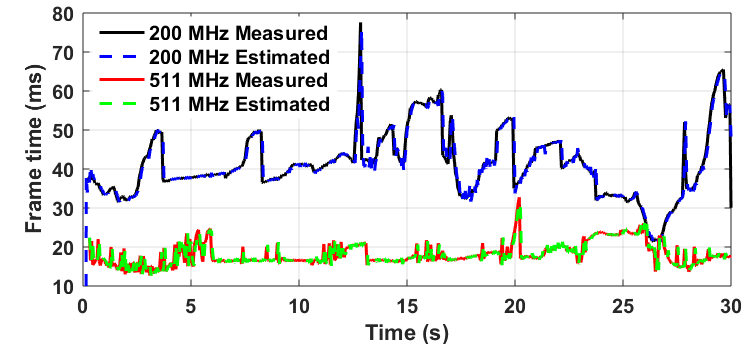}
    \vspace{-5mm}
	\caption{Frame time prediction while running Nenamark2 benchmark on Minnowboard MAX platform~\cite{IntelMinnowboard} with Android OS.}
	\label{fig:frame_time}
\vspace{-3.5mm}
\end{figure}

Dynamic resource management techniques can utilize the sensitivity of performance and power consumption to controlled parameters, such as the operating frequency, to meet the design objectives. 
Auto-regressive models~\cite{dietrich2014lightweight} and recursive least squares (RLS)-based performance models with exponential forgetting factor~\cite{gupta2018online} are proposed to guide DRM algorithms. 
A PID-controller based workload prediction model for graphics workload is presented in~\cite{dietrich2010lms}.
To maintain a high accuracy under different applications, these models need to utilize performance counters that capture workload characteristics and adapt to dynamic changes at runtime.
To this end, an online feature selection and adaptive learning technique using RLS for modeling the performance of integrated GPUs is presented in~\cite{gupta2018staff}.
Figure~\ref{fig:frame_time} shows the frame processing time for the Nenamark2 benchmark.
The estimated frame time closely follows the measured value at different operating frequencies with less than 5\% error.
The authors also show that this online learning technique can be applied to power consumption modeling and performance modeling of CPU cores~\cite{bhat2018online}.

\subsection{Performance Models for NoCs} \label{sec:noc_model}

NoC performance models are useful for fast design space exploration and speeding-up full system simulations~\cite{kiasari2013analytical, qian2015support}.
State-of-the-art techniques view the NoC as a network of queues and construct performance models using queuing theory~\cite{mandal2019analytical,ogras2013modeling}.
However, analytical NoC performance models depend on the target architecture and cannot be easily generalized across multiple NoC configurations.
To increase robustness of the NoC models, a machine learning based performance analysis technique has been proposed~\cite{qian2015support}. 
In this technique, the channel and source waiting times for the NoC are estimated through analytical models.
Then, the waiting time obtained from the analytical models and the waiting time obtained from an NoC simulator are used as features to learn support vector regression (SVR)-based model to estimate NoC performance.
This technique relies on the model constructed offline similar to other existing NoC performance analysis approaches. 
These offline techniques have been utilized successfully for design space exploration due to their steady-state accuracy. 
However, they are not designed to adapt to rapidly changing workload characteristics at runtime. 
Therefore, there is a need for online techniques that can learn or adapt the NoC performance models to dynamically varying workloads, 
similar to the models described in Section~\ref{sec:gpu_model}.



%% file: files/control.tex
\section{Dynamic Control and Resource Management}\label{sec:control}
This section describes the use of the predictive models for DRM using two modern approaches: imitation learning (IL) and an explicit nonlinear model predictive control (NMPC).

\subsection{Application of Imitation Learning Methodology to DRM}
\label{sec:ml_framework}






We first describe an offline IL methodology to construct resource management policies that can be used to bootstrap the online learning process. Afterward, we describe an RL methodology for online policy learning and discuss its critical drawbacks. Finally, we present a model-guided online IL methodology to overcome the drawbacks of RL.

\subsubsection{Imitation Learning} \label{sec:offline_IL}


IL is a supervised learning technique, which is used to obtain an optimal solution for sequential decision making problems.
An IL policy follows demonstrations shown by an expert which is also known as Oracle~\cite{schaal1999imitation}.

In the power management domain, the Oracle is typically constructed offline 
by executing a set of known applications on the target platform 
while recording the relevant system states~\cite{kim2017imitation}.
The system states are then used to construct an Oracle policy optimizing certain metrics.
For example, the offline IL policy proposed in~\cite{mandal2019dynamic} first segments the applications into workload-conservative snippets, i.e. each snippet has a fixed number of instructions~\cite{gupta2017dypo}.
Then, each snippet in the set of target applications is executed at each configuration supported by the SoC.
The system states shown in Table~\ref{tab:counters} are recorded during each execution.
Finally, these system states and power consumption measurements are used to construct Oracle policies 
which optimize different objectives (e.g., energy consumption, performance-per-watt). 
Construction of Oracle policies that enable optimal sequences of control decisions can involve the use of dynamic programming or other optimization techniques.

Oracle policies cannot be used directly at runtime due to excessive storage requirements.
Likewise, they cannot be constructed at runtime due to high computational complexity.
Therefore, there is a need for a policy, 
which can well approximate the Oracle and provide the power management decisions at runtime.
To construct the online policy, 
any off-the-shelf machine learning models can be used: linear regression and regression tree-based models are used in~\cite{kim2017imitation, mandal2019dynamic}.


\begin{table}[t]
	\centering
	\caption{Data collected in each snippet}
 	\vspace{-7mm}
    \label{tab:counters}
	\begin{tabular}{@{}ll@{}}
		\\ \midrule
		Instructions Retired             & Noncache External Memory Request      \\
		CPU Cycles                       & Total Little Cluster Utilization    \\
		Branch Miss Prediction           & Per Core Big Cluster Utilization \\
		Level 2 Cache Misses             & Total Chip Power Consumption      \\
		Data Memory Access               &               \\ \bottomrule
	\end{tabular}
\end{table}


IL policies designed offline perform well under the workloads 
used during training. 
However, their performance may drop for new applications seen at runtime 
since the policies may not generalize well.
This is illustrated by a study on Odroid-XU3 platform 
with two big.LITTLE CPU clusters, 
whose frequencies can be controlled independently~\cite{mandal2019dynamic}. 
An IL policy was trained on applications from Mi-Bench~\cite{guthaus2001mibench} suite.
Table~\ref{tab:act_accur} shows 
that the IL policy performs well on the training applications from Mi-Bench suite.
However, the energy significantly increases when compared to the Oracle
policy for applications from Cortex~\cite{thomas2014cortexsuite} and PARSEC~\cite{bienia2008parsec} benchmark suites. 
Hence, there is a need for adaptive learning techniques for dynamic resource management. 

\vspace{1mm}
\subsubsection{Reinforcement Learning} \label{sec:rl}

Reinforcement learning is a well explored model-free online learning technique where the policy is learned through the reward provided by the environment.
There are two well-known approaches to perform RL: table-based and deep Q-learning based.
Table-based RL technique is not practical due to the large storage requirement.
Deep Q-learning based RL approach is also not suitable for runtime resource management of SoCs. 
There are two major drawbacks for RL-based approach.
\begin{itemize}
    \item RL-based controller learns through the reward obtained from the environment. Hence, the efficiency of the controller critically depends on the amount of exploration performed by the controller. This behavior makes the RL controller to take a long time to converge to an optimal policy that is not suitable for SoC control since the workload characteristics can quickly change at runtime.
    \item Designing a good reward function is critical in RL-based learning technique, since reward function drives the learning quality of the controller. However, there is no systematic way to construct the reward function. 
\end{itemize}
To address the challenges posed by RL techniques, we can apply an online IL methodology as described next.

\begin{table}[t]
\centering
\caption{Energy (normalized w.r.t. the Oracle) for applications from Mi-Bench, Cortex and PARSEC suite with an IL policy trained with applications from Mi-Bench suite.}
\label{tab:ml_accuracies}
\setlength\tabcolsep{2pt}
\label{tab:act_accur}
\footnotesize{
\begin{tabular}{@{}c|c|c|c|c|c|c|c|c|c@{}}
\toprule
                                                                 & \multicolumn{4}{c|}{Mi-Bench} & \multicolumn{3}{c|}{Cortex} & \multicolumn{2}{c}{PARSEC}                                                                                   \\ \midrule
                                                                 & BML   & Djkstr & FFT  & Qsort & MtnEst   & Spctrl   & Kmns  & \begin{tabular}[c]{@{}c@{}}Blkschls\\ 2T\end{tabular} & \begin{tabular}[c]{@{}c@{}}Blkschls\\ 4T\end{tabular} \\ \midrule
\begin{tabular}[c]{@{}c@{}}Normalized\\ Energy\end{tabular} & 1.00  & 1.01   & 1.00 & 1.00  & 1.13     & 1.09     & 1.76  & 1.86                                                  & 1.47                                                  \\ \bottomrule
\end{tabular}
}
\vspace{-6mm}
\end{table}

\vspace{1mm}
\subsubsection{Model-Guided Online Imitation Learning} \label{sec:online_IL}

In general, providing online supervision on the best configuration to update the policy is a challenging problem.
This challenge is addressed by a recent online-IL technique~\cite{mandal2020energy} using predictive power and performance models, such as those presented in Section~\ref{sec:ana_model}.
This technique starts with an IL policy, and predictive power and performance models, 
constructed offline, e.g., using the data available at design time. 
At runtime, the system state data
listed in Table~\ref{tab:counters} are collected at the end of each snippet 
of the application (as described in Section~\ref{sec:offline_IL}) and are 
used to continuously update the power and performance models (as explained in Section~\ref{sec:gpu_model}).
Before each control decision, these models and the state data are used to estimate the energy consumption of candidate configurations in a local neighborhood of the current configuration.
Since this computation introduces power and performance overheads, 
hardware counter values observed at runtime for the current configuration can be reused to approximate the power consumption of other configurations. Note that in general, the hardware counters do not remain the same across different configurations. Predicting changes in counters is a complex open problem, which involves modeling of system dynamics and the state transition probability.
Finally, the configuration with the minimum energy consumption is marked as the optimal configuration 
and added to the runtime approximation of the Oracle. 

The last step in the online-IL methodology is to incrementally update the parameters of the policy as a function of the workload.  
The best configuration found by the analytical models (i.e., the runtime approximation of the Oracle policy) and performance counters in Table~\ref{tab:counters} are inserted in a buffer after each policy decision.
This training data is aggregated until the buffer is full. 
Subsequently, the  policy is updated using the training data and the buffer is reset.
In our particular setting, the policy is represented as a neural network and 
it is updated using the back-propagation algorithm~\cite{hecht1992theory}. 
The size of this buffer determines the training accuracy and implementation overhead.
Experimental evaluation in~\cite{mandal2020energy} shows that input and output control state for 100 epochs provides close to 100\% accuracy in adapting to new applications. 
The corresponding storage overhead for this buffer is less than 20KB. 

\begin{figure}[h]
	\centering
	\vspace{-3mm}
	\includegraphics[width=1\linewidth]{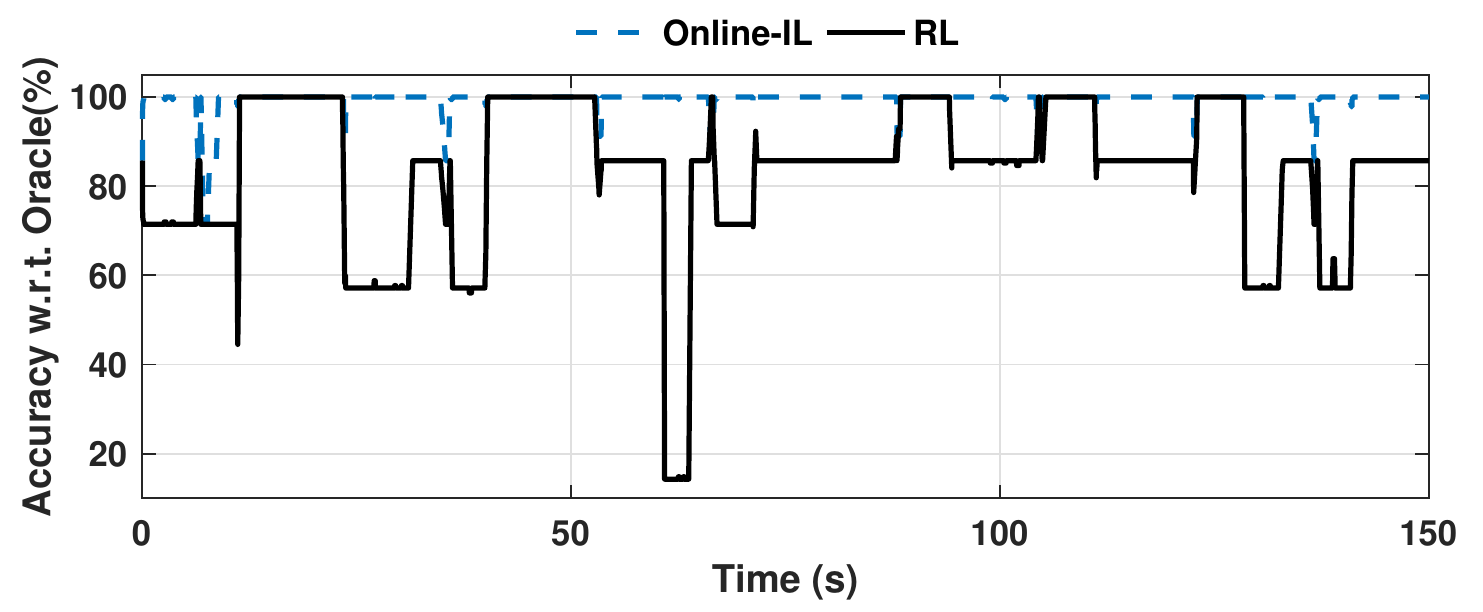}
    \vspace{-6mm}
	\caption{Comparison of convergence for the frequency of big cores with a sequence of applications executed on ODROID-XU3 platform.}
	\vspace{-3mm}
	\label{fig:convergence}
\end{figure}
\begin{figure}[t]
	\centering
	\vspace{-2mm}
	\includegraphics{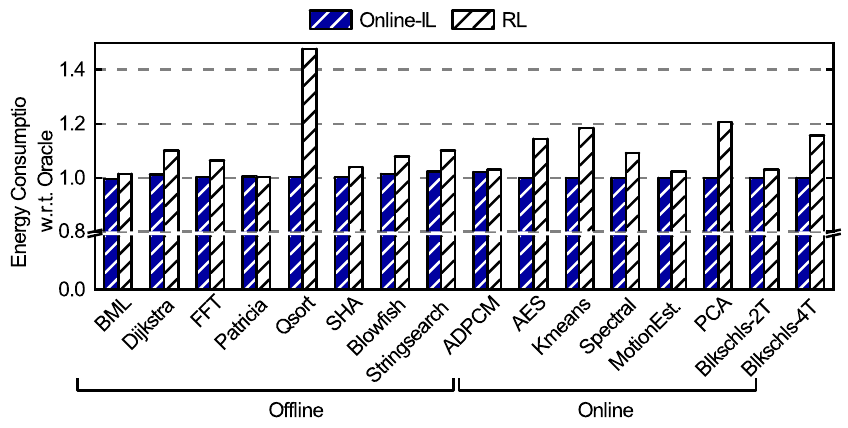}
    \vspace{-4mm}
	\caption{Energy comparison between proposed IL and RL approach. Both models are trained offline with the same applications. The offline model is updated online with a sequence of applications labeled `online'.}
	\label{fig:energy}
	\vspace{-6mm}
\end{figure}

Figure~\ref{fig:convergence} compares the proposed online IL policy and a policy found using RL. 
Both policies are trained offline with Mi-Bench applications.
Then, the initial policies are adapted while running a sequence of applications from Cortex and PARSEC.
The IL policy converges to the Oracle policy (close to 100\% accuracy) within 6 seconds, which is only 4\% of the total execution time.
In contrast, the RL policy does not converge even after the whole sequence.
Furthermore,  
Figure~\ref{fig:energy} shows that the energy consumption using the IL policy are similar to the results with Oracle policy
for all benchmark applications. 
In contrast, the energy consumption achieved with the RL policy is up to 1.4$\times$ higher than the Oracle policy. 

\subsection{Explicit Nonlinear Model Predictive Control for Multi-variable Power Management}\label{sec:nmpc}

Power and thermal management of modern SoCs involve multiple output control variables that may have different control overhead. For instance, integrated GPUs have the ability to manage operating frequency and voltage as well as power gating individual GPU slices.  For deadline-driven graphics applications the control decisions should be done in a coordinated fashion to maximize energy efficiency while meeting the target frames per second (FPS). A multi-rate control is generally required to  handle the differences in the time granularity of the control knobs: e.g., changing the number of active slices takes significantly longer time and requires more energy than changing the frequency and voltage values. Moreover, the implementation of this technique requires very low complexity, runtime and memory overhead since the control algorithms are implemented in firmware and hardware.

A novel multi-rate predictive control algorithm has been proposed to manage DVFS and the number of active slices within the GPU in a coordinated fashion~\cite{mercati_2017}. 
The formulation utilizes predictive sensitivity models for the control knobs to abstract
the underlying system details.
These sensitivity models can be learned off- or on-line and take input from a small subset of the available performance counters as explained in Section~\ref{sec:gpu_model}. The multi-rate controller is comprised of two sub-controllers as follows:
\textit{\textbf{Slow-rate controller}} orchestrates the dynamic assignments of the operating frequency (DVFS) and the number of active slices at a coarse time granularity. 
\textit{\textbf{Fast-rate controller}} manages the operating frequency (DVFS) at a finer time granularity using hardware support for fast changes in frequency and voltage. 
It applies a state-space control since it is known to be robust for handling discrete control problems. The formulation shows that this is a non-linear constrained control that can be optimally solved via nonlinear model predictive control (NMPC)~\cite{allgower2012nonlinear}. 
However, NMPC is not suitable for low-overhead runtime applications due to its high computational overhead.
To overcome this challenge, explicit NMPC is proposed.
Explicit NMPC addresses the constrained non-linear control problem with low overhead~\cite{chakrabarty_2017, zhang_2016}.
A recent technique applies explicit NMPC in the domain of dynamic resource management~\cite{mercati_2017}.
This technique approximates the surface of NMPC control using simple 
regression models that can be implemented with low overhead while achieving near optimal control.

\begin{figure}[t]
	\centering
	    \includegraphics[width=1\linewidth]{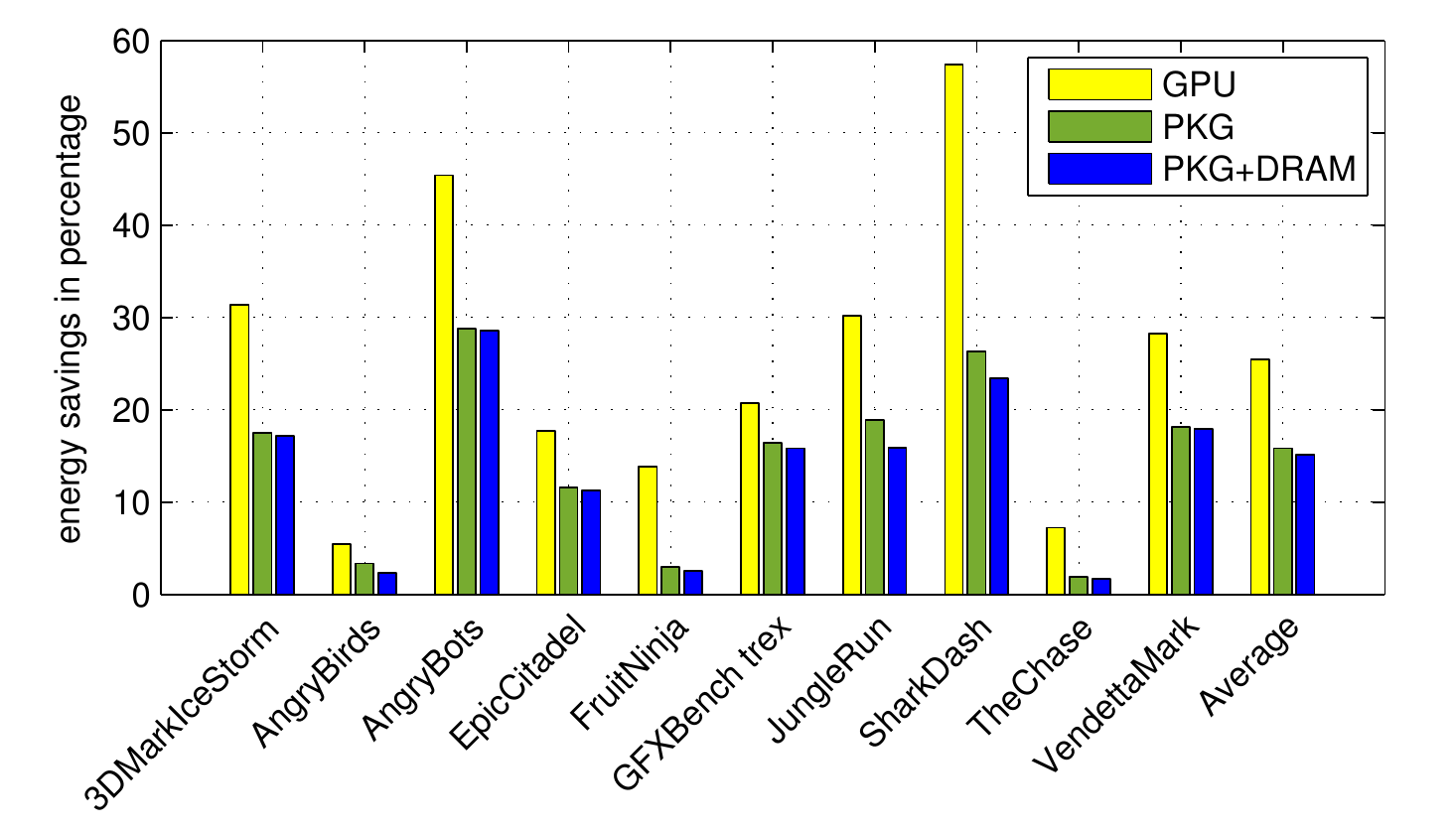}
    \vspace{-6mm}
	\caption{Energy savings of the explicit NMPC algorithm compared to baseline power management.}
	\label{fig:nmpc_res_energy}
	\vspace{-5mm}
\end{figure}

Figure~\ref{fig:nmpc_res_energy} shows the normalized energy savings of the explicit NMPC algorithm with respect to the baseline algorithm on the Intel\textregistered\  Core\texttrademark\  i5 platform. It reports the savings for the GPU alone, for the system package (PKG) and for the package and memory subsystem (PKG+DRAM). For the GPU, the savings range from 5\% in the case of Angrybirds to 58\% for SharkDash applications, while the average is 25\%. This technique achieves savings of approximately 15\% for both the PKG and the PKG+DRAM cases. The energy savings are consistent at different platform thermal conditions and are achieved with a negligible performance overhead of 0.4\%.

%% file: files/conclusion.tex
\section{Conclusions}

Runtime resource management of heterogeneous SoCs is crucial to maintain a trade-off between power consumption and performance. 
To this end, we presented state-of-the-art
online learning algorithms for modeling design objectives and a model-guided dynamic resource management.
We also
demonstrated applications of explicit nonlinear model predictive control and imitation learning for dynamic resource management of heterogeneous SoCs. 

Multiple research problems remain open, including low-cost implementation of IL and RL suitable for firmware implementation, and generalization of explicit model predictive control to broader class of systems and higher dimensional spaces of input and output control variables.